\documentclass[cameraready]{Interspeech}
 
\title{Pruning as Regularization: Sensitivity-Aware One-Shot Pruning in ASR}
 
\author[affiliation={1}]{Julian}{Irigoyen}
\author[affiliation={2}]{Arthur}{S\"ohler}
\author[affiliation={3,\dagger}]{Andreas S\o eborg}{Kirkedal}
 
\address{
    $^1$ Danske Bank, Copenhagen, Denmark \\
    $^2$ Copenhagen Business School, Copenhagen, Denmark \\
    $^3$ Boost.ai, Copenhagen, Denmark
}
 
\email{jiri@danskebank.dk, arso21ab@student.cbs.dk, andreas.kirkedal@boost.ai}
 
 
\newcommand{\Whisper}{Whisper}

\usepackage{amsmath,amssymb,amsfonts}
\usepackage{graphicx}
\usepackage{booktabs}
\usepackage{multirow}
 
\begin{document}

\maketitle
\def\thefootnote{$\dagger$}\footnotetext{Work done while at Jabra (GN Group).}
\def\thefootnote{\arabic{footnote}} 
\begin{abstract}
Neural network pruning is typically framed as a post-training compression technique.
We show that for encoder-decoder ASR Transformers, one-shot magnitude pruning can act
as a strong implicit regularizer: removing redundant parameters improves generalization
without fine-tuning. Using Whisper-small, we introduce a sensitivity diagnostic
combining gradient and Fisher criteria to identify pruning-fragile vs.
pruning-resilient components. This reveals an encoder-decoder asymmetry: decoder FFNs
are pruning-fragile, whereas decoder self-attention and late encoder layers contain
removable redundancy. Without fine-tuning, pruning 50\% of decoder self-attention
improves WER by 2.38\% absolute on LibriSpeech test-other; pruning the last four
encoder layers at 50\% yields 1.72\% improvement. Gains persist on Common Voice and
TED-LIUM, demonstrating cross-corpus generalization. At 40.8\% sparsity,
sensitivity-aware compression preserves near-baseline accuracy where global magnitude
pruning collapses.
\end{abstract}

\section{Introduction}
\label{sec:intro}

Over-parameterization in neural networks facilitates training convergence but often
introduces redundancies that can impair generalization. Conventional regularizers such
as dropout \cite{srivastava2014dropout}, weight decay \cite{krogh1991simple}, and batch
normalization \cite{ioffe2015batch} mitigate this through stochastic masking or
parametric constraints. Beyond these established techniques, we demonstrate that
architectural analysis can reveal specific over-parameterized components where pruning
acts as an implicit regularizer.

While pruning has traditionally been studied as a compression technique aimed at
reducing computational
cost~\cite{han2015learning,lecun1989optimal,hassibi1993optimal}, we provide empirical
evidence that \emph{post-training one-shot magnitude pruning} without fine-tuning can
\emph{improve} WER for a strong foundation ASR model (\Whisper{}-small), if
strategically allocated based on component sensitivity. We identify redundant
parameters using gradient and Fisher-based sensitivity diagnostics, and validate them
through component- and layer-wise pruning experiments. These diagnostics enable
targeted pruning that can yield meaningful performance gains and improve robustness to
unseen acoustic conditions. They can also guide aggressive one-shot pruning and surpass
standard pruning methods that collapse at similar sparsity levels.

Our contributions are threefold: (1) \textbf{Sensitivity diagnostics framework:} A
replicable methodology that combines gradient-based and Fisher Information-based
sensitivity analysis to quantify per-component and per-layer tolerance to one-shot
magnitude pruning in ASR systems. (2) \textbf{Pruning-as-regularization:} Empirical
demonstration that targeted pruning of identified redundant components improves ASR
generalization across acoustically diverse corpora (LibriSpeech, Common Voice,
TED-LIUM) without any fine-tuning. (3) \textbf{Sensitivity-aware compression:} Our
diagnostics enable safer one-shot compression via component-specific sparsity
allocation, preserving accuracy even at high sparsity.

The proposed methodology is model-agnostic and can be applied to other Transformer-based ASR architectures beyond \Whisper{}, as it relies on post-training gradients and
sensitivity estimates rather than model-specific characteristics. The resulting sensitivity profiles must
be validated separately for each architecture.

\section{Related Work}
\label{sec:prior}

\textbf{Pruning for compression.} Researchers originally introduced pruning to reduce
model size and computation
cost~\cite{lecun1989optimal,hassibi1993optimal,han2015learning}. Architectural analyses
in Transformers reveal redundancy in attention
heads~\cite{voita2019analyzing,michel2019heads,behnke2020losing}, with
surveys~\cite{gale2019state} showing magnitude pruning matching sophisticated approaches
in terms of compression-accuracy trade-offs. Our focus differs: we leverage magnitude
pruning primarily as a \emph{regularizer}.

\textbf{Pruning as regularization.} Prior studies link pruning to generalization.
Bartoldson et al.~\cite{bartoldson2020generalization} link immediate accuracy drops to
larger eventual improvements with continued training, while Jin
et al.~\cite{jin2022pruning} attribute improvements to a combination of longer training
and reduced model size. We extend this perspective to \emph{post-training} one-shot
pruning in ASR, reporting improvements without any fine-tuning.

\textbf{One-shot and sensitivity-based pruning.} The Lottery Ticket
Hypothesis~\cite{frankle2019lottery} showed sparse subnetworks in dense models,
inspiring one-shot methods like SNIP~\cite{lee2019snip}, GraSP~\cite{wang2020picking},
and SynFlow~\cite{tanaka2020pruning} that identify masks at initialization via
gradient/flow criteria. Second-order approaches
(WoodFisher~\cite{singh2020woodfisher}, Group Fisher~\cite{liu2021group}) drive pruning
decisions by approximating the Hessian. Kwon et al.~\cite{kwon2022fast} introduced
post-training Fisher-based pruning without retraining. We extend these approaches by
applying both \emph{first- and second-order} analyses post-training to diagnose
regularization opportunities.

\textbf{ASR pruning.} Prior ASR pruning research primarily emphasizes compression
objectives. Peng et al.~\cite{peng2023structured} explored structured pruning for
efficiency, Kim et al.~\cite{kim2023adapt} investigated adaptive multilingual pruning
strategies, and Gu et al.~\cite{gu2024sparseWAV} developed sparse ASR models. In
contrast, we demonstrate that one-shot magnitude pruning can \emph{improve accuracy}
without fine-tuning, revealing pruning's dual role as both a compression tool and a
regularization technique.

\section{Methodology}
\label{sec:method}

\subsection{Theoretical foundation: pruning as regularizer}
\label{ssec:theory}

Classical regularization augments the empirical risk with explicit penalties like
$\lVert\theta\rVert_2^2$ (weight decay) or $\lVert\theta\rVert_1$
(lasso)~\cite{krogh1991simple,tibshirani1996regression}. Magnitude pruning instead acts
as \textit{implicit regularization} by directly removing parameters rather than
penalizing them.

For a neural network with parameters $\theta \in \mathbb{R}^d$ and target sparsity
level $\rho \in [0,1)$, one-shot magnitude pruning retains the $(1-\rho)d$ parameters
with largest absolute values:
\begin{equation}
  \theta_\mathcal{M} = \{\theta_i : i \in \mathcal{M}\}, \quad
  |\mathcal{M}| = (1-\rho)d,
\end{equation}
where $\mathcal{M} \subset \{1, \dots, d\}$ indexes the retained parameters. The
pruned network $f_{\theta_\mathcal{M}}$ operates on this sparse subset.

This deterministic capacity constraint differs fundamentally from stochastic regularizers like dropout: by eliminating low-magnitude parameters and preserving dominant ones, pruning can improve generalization in over-parameterized networks.

\subsection{Sensitivity analysis framework}
\label{ssec:sensitivity}

We assess pruning robustness at both module and layer granularity within both the
encoder ($\mathcal{E}$) and decoder ($\mathcal{D}$) components. Our framework combines
three complementary approaches: (i) first-order (gradient-based) sensitivity,
(ii) second-order (curvature-based) sensitivity using Fisher Information, and
(iii) empirical component-wise pruning validation. These diagnostics enable
component-specific sparsity allocation that avoids fragile components and concentrates
sparsity in resilient regions.

\textbf{First-order sensitivity.}
Following SNIP~\cite{lee2019snip}, for module
$m \in \{\mathcal{E}, \mathcal{D}\}$ with parameters $\theta_m$ we compute a
normalized gradient-based sensitivity score:
\begin{equation}
  \label{eq:firstorder}
  S_g^{(m)} := \frac{1}{N}\sum_{i=1}^{N}
    \frac{\lVert\nabla_{\theta_m}\mathcal{L}(x_i,y_i)\rVert_2}
         {\lVert\theta_m\rVert_2},
\end{equation}
where $\mathcal{L}(x_i, y_i; \theta)$ denotes the negative log-likelihood
(cross-entropy) loss for acoustic input $x_i$ and transcription $y_i$, evaluated over
$N$ samples from the validation set. This scale-normalized metric captures the
immediate sensitivity of the loss to small perturbations in $\theta_m$. Higher values
indicate greater pruning fragility.

\textbf{Second-order sensitivity.}
To capture loss curvature, we approximate the diagonal of the Fisher Information Matrix
(FIM). For negative log-likelihood losses (e.g., cross-entropy), the FIM diagonal can
be estimated as~\cite{kirkpatrick2016overcoming}:
\begin{equation}
  \label{eq:fim}
  F_{jj} \approx \frac{1}{N}\sum_{i=1}^{N}
    \left(\nabla_{\theta_j}\mathcal{L}(x_i,y_i;\theta)\right)^2,
\end{equation}
for parameter index $j \in \mathcal{I}_m$, where $\mathcal{I}_m$ denotes the parameter
indices belonging to module $m$. The module-level Fisher sensitivity is then:
\begin{equation}
  \label{eq:modulefim}
  S_h^{(m)} := \frac{1}{|\mathcal{I}_m|}\sum_{j \in \mathcal{I}_m} F_{jj}.
\end{equation}
Large $S_h^{(m)}$ values indicate high local curvature (parameters are tightly
constrained by the loss landscape), suggesting lower tolerance to pruning. Conversely,
low values suggest flat regions where pruning may be safer.

\textbf{Empirical pruning sensitivity.}
Beyond these analytic diagnostics, we empirically assess pruning sensitivity by
measuring performance degradation (or improvement) after pruning. Components
maintaining or improving performance when pruned reveal implicit regularization
opportunities.

We employ one-shot, unstructured magnitude pruning. For each architectural component, we rank all weights by absolute value and zero out the lowest-magnitude fraction corresponding to the target sparsity $\rho$. All other model parameters remain unchanged. We evaluate each component independently in both encoder and decoder, as well as layer subsets (early layers 1--4, middle 5--8, late 9--12).

For reproducibility, given target sparsity $\rho$ and component $c$, we compute the
threshold $\tau(\rho, c)$ as the $(100\rho)$-th percentile of
$\{|\theta_j| : j \in c\}$, then set $\theta_j \leftarrow 0$ for all $j$ where
$|\theta_j| \leq \tau(\rho, c)$.

\subsection{Experimental setup}
\label{ssec:data}

\textbf{Model.} We use the publicly released \Whisper{}-small
checkpoint~\cite{radford2022whisper}, a 244M-parameter encoder--decoder Transformer
with 12 encoder and 12 decoder layers. All evaluations employ \Whisper{}'s standard beam
search decoding pipeline.

\textbf{Datasets and evaluation protocol.} Primary evaluation uses
LibriSpeech~\cite{panayotov2015librispeech} \textit{test-clean} (clean speech) and
\textit{test-other} (challenging acoustic conditions). We treat \textit{test-other} as
our primary reporting metric due to its greater acoustic variability.

To assess cross-corpus generalization, we validate on 2000 randomly sampled utterances
from Common Voice v15 English (crowd-sourced recordings with diverse accents and
conditions~\cite{ardila2020common}) and TED-LIUM Release~3 (conference talks with
natural reverberation and spontaneous speech~\cite{hernandez2018ted}).

\textbf{Metrics.} We quantify ASR performance using word error rate (WER) and
character error rate (CER). Pruning resilience is measured through $\Delta$WER on
\textit{test-other} relative to our unpruned baseline (11.64\%): negative values
indicate improvement. This baseline reflects our normalization and scoring pipeline rather than an attempt to reproduce the WER reported by Radford et al.~\cite{radford2022whisper}; conclusions rely on our own paired deltas rather than external WER values. Note that unstructured sparsity masks are not exploited by standard dense inference kernels in our implementation, hence real-time factor (RTF) remains unchanged.

\textbf{Evaluation procedure.} Our sensitivity diagnostics ($S_g$, $S_h$) are computed
post-training using a calibration subset from LibriSpeech validation data, purely for
\emph{analysis} (no parameter updates occur). Sparsity levels are swept in 10\%
increments from 0\% to failure. Due to the number of tests and space constraints, we
report salient sparsity points that best illustrate regularization gains or failure
modes. Pruning configurations identified on LibriSpeech are then evaluated
\emph{unchanged} (same masks, no recomputation) on Common Voice and TED-LIUM to
demonstrate cross-dataset generalization. We never use target dataset labels to inform
pruning decisions.

\textbf{Computing infrastructure.} Experiments were performed on HPC nodes with NVIDIA
H100 GPUs. Sensitivity diagnostics (gradient and Fisher passes) take approximately
11 minutes. WER evaluations across all sparsity levels take approximately
15 minutes per component. No model training was performed.

\section{Experimental Results \& Implications}
\label{sec:results}

\subsection{First- and second-order sensitivity diagnostics}
\label{ssec:diag-sensitivity}

\begin{figure}[t]
  \centering
  \includegraphics[width=\linewidth]{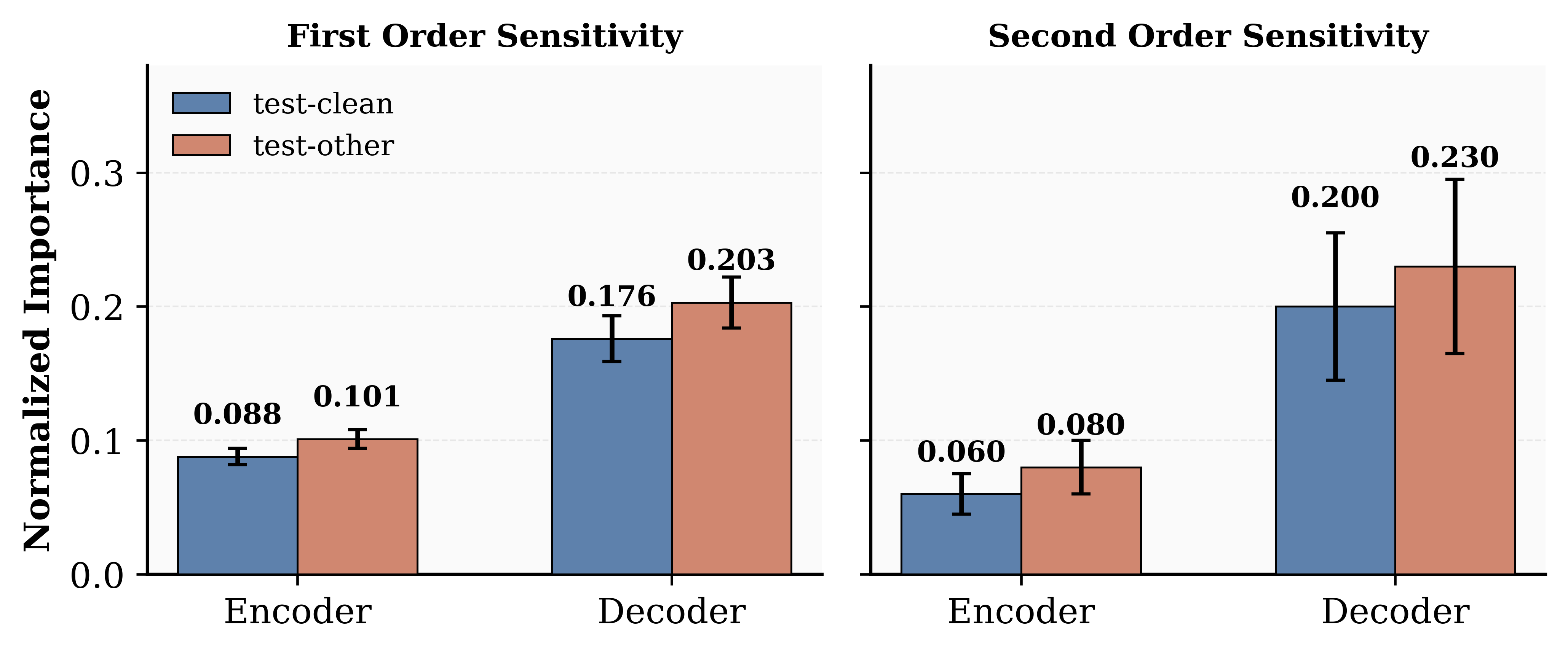}
  \caption{First-order (gradient) and second-order (Fisher) sensitivity for
    encoder/decoder on LibriSpeech \textit{dev-clean} and \textit{dev-other}
    validation subsets. Higher values indicate greater pruning sensitivity.}
  \label{fig:sensitivity_test}
\end{figure}

Figure~\ref{fig:sensitivity_test} reports module-level sensitivity for encoder and
decoder using both first-order (gradient) and second-order (Fisher) criteria. Across
both diagnostics, the decoder exhibits consistently higher sensitivity values, which
indicate greater vulnerability to magnitude pruning. These results indicate an
encoder--decoder asymmetry in pruning robustness that we verify empirically and discuss
below.

\subsection{Architectural Encoder--Decoder Asymmetry}
\label{ssec:encoder-decoder}

\begin{figure}[t]
  \centering
  \includegraphics[width=\linewidth]{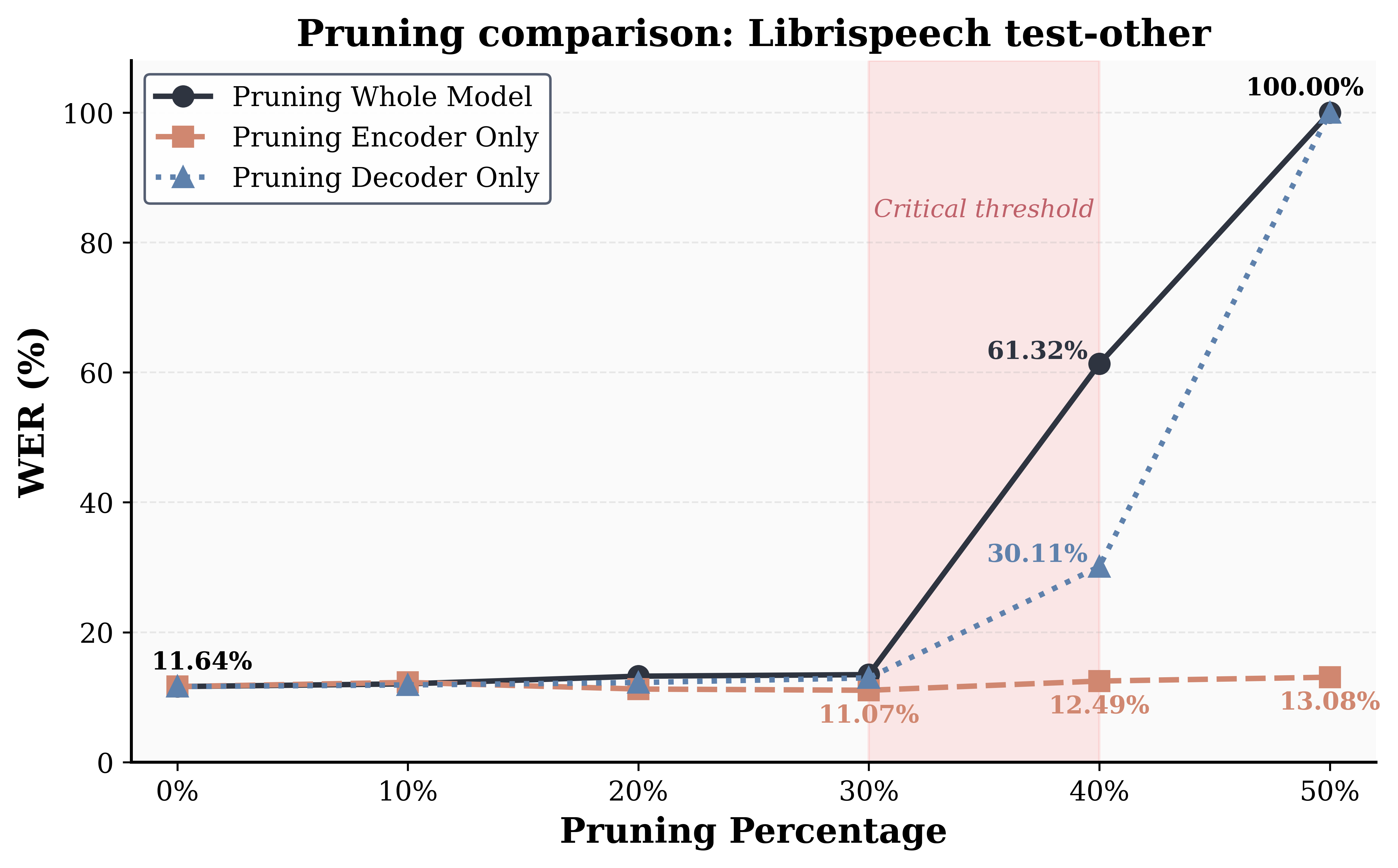}
  \caption{WER vs.\ sparsity when pruning the whole model vs.\ encoder/decoder in
    isolation on LibriSpeech \textit{test-other}.}
  \label{fig:pruning_comparison}
\end{figure}

Building upon the previous diagnostics, Figure~\ref{fig:pruning_comparison} reports
WER on \textit{test-other} when pruning (i) the entire model (global magnitude
pruning), (ii) the encoder only, or (iii) the decoder only. Global pruning exhibits
catastrophic failure between 30--40\% sparsity, confirming that naive uniform sparsity
allocation is unsuitable. Separate encoder/decoder pruning reveals the decoder drives
this collapse, yielding 30.11\% WER at 40\% sparsity, whereas encoder-only pruning
remains relatively close to baseline (12.49\%).

This asymmetry is counterintuitive: the encoder comprises 36.47\% of model parameters
while the decoder accounts for 63.53\%. Despite having fewer parameters, the encoder
tolerates pruning significantly better than the more heavily parameterized decoder.
Notably, encoder-only pruning at 30\% sparsity \emph{improves} over baseline
($\Delta$WER\,=\,--0.57\%), providing initial evidence for selective pruning inducing
beneficial regularization.

\subsection{Layer-wise granularity}
\label{ssec:layer}

We further analyze layer-level sensitivity by dividing encoder and decoder into three
blocks: early (layers 1--4), middle (5--8), and late (9--12).
Table~\ref{tab:layer_pruning} shows results at 50\% sparsity.

\begin{table}[th]
  \caption{WER (\%) for layer-wise pruning at 50\% sparsity.}
  \label{tab:layer_pruning}
  \centering
  \small
  \begin{tabular}{lrrr}
    \toprule
    Layers        & Clean & Other & $\Delta$Other \\
    \midrule
    \multicolumn{4}{c}{\textit{Encoder}} \\
    \midrule
    Early (1--4)  & 3.43  & 12.89 & +1.25           \\
    Middle (5--8) & 3.62  & 11.84 & +0.20           \\
    Late (9--12)  & \textbf{3.31} & \textbf{9.92} & \textbf{--1.72} \\
    \midrule
    \multicolumn{4}{c}{\textit{Decoder}} \\
    \midrule
    Early (1--4)  & 38.68 & 79.14 & +67.50          \\
    Middle (5--8) & 3.37  & 12.63 & +0.99           \\
    Late (9--12)  & 6.77  & 20.54 & +8.90           \\
    \bottomrule
  \end{tabular}
\end{table}

Encoder parameter importance decreases with depth: early layers exhibit moderate
fragility (+1.25\% $\Delta$WER), likely due to their role in low-level acoustic
feature extraction from the speech signal. Late encoder layers (9--12) not only tolerate 50\% pruning but
actively \emph{improve} performance by 1.72\% absolute WER (14.8\% relative
improvement on \textit{test-other}). Their gains suggest they may overfit to training-data acoustic patterns, and pruning acts as a corrective.

The decoder shows a U-shaped vulnerability profile: early layers (conditioning on
encoder context) and late layers (producing final token distributions) are critical,
while middle layers show tolerance. Early decoder layers are extremely fragile
(+67.50\% at 50\%), suggesting essential cross-attention computations that cannot be
compressed without fine-tuning.

We attribute this asymmetry to the functional roles of each component. The decoder's autoregressive generation process is tightly constrained by the loss: pruning-induced perturbations compound across output tokens, with errors in early decoder blocks propagating and amplifying at every subsequent autoregressive step. The encoder, by contrast, processes the full acoustic sequence in parallel through non-causal self-attention and residual pathways, distributing information across multiple heads, layers, and time steps. This overlap provides a natural redundancy buffer. Consequently, upper encoder blocks act as incremental refinements on an already stable representation, making them safe targets for pruning, while early decoder blocks shape the state reused at every generation step and cannot be disturbed without fine-tuning.

\subsection{Component-specific pruning}
\label{ssec:component}

Table~\ref{tab:component_results} summarizes salient sparsity points in our
component-level pruning sweeps, revealing asymmetries in regularization capacities
across elements. Encoder \textbf{Feed-forward networks (FFNs)} tolerate 40\% pruning
with slight improvements (--0.52\%), while decoder FFNs catastrophically fail beyond
30\% (+40.99\% at 40\%). This aligns with our sensitivity diagnostics and confirms
decoder FFNs are critical for maintaining output quality. \textbf{Attention mechanisms}
exhibit the opposite pattern. Encoder self-attention is fragile (fails above 50\%),
whereas decoder self-attention tolerates aggressive pruning and delivers the largest
regularization gain: 50\% sparsity yields --2.38\% WER (20.4\% relative improvement).
This persists even at 60\% sparsity (--1.43\%). We hypothesize this occurs because
pruning eliminates redundant or conflicting attention heads, resulting in more focused
attention patterns. For the auxiliary components, modest pruning of \textbf{LayerNorms}
(10\%, --0.63\%) and the \textbf{final output projection} (20\%, --1.10\%) yields
consistent gains. \textbf{Bias parameters} are highly sensitive (+7.53\% at just
10\%), as they often carry crucial offset information. Finally, front-end
\textbf{Convolutional layers} used for initial feature extraction tolerate 50\% pruning
with no WER degradation, suggesting greater importance for training stability and convergence than as a limiting factor for inference performance. Pruning eliminates redundant local filters compensated by downstream learned pathways.

\begin{table}[th]
  \caption{Component-wise pruning results. WER (\%) at varying sparsity.}
  \label{tab:component_results}
  \centering
  \begin{tabular}{lcrrr}
    \toprule
    Component      & Sparsity & Clean & Other & $\Delta$Other   \\
    \midrule
    Baseline       &  0 & 3.45  & 11.64 & --              \\
    \midrule
    Enc Self-Attn  & 50 & 3.91  & 13.06 & +1.42           \\
                   & 60 & 6.06  & 19.72 & +8.08           \\
    Enc FFNs       & 40 & 3.46  & 11.04 & \textbf{--0.52} \\
    \midrule
    Dec Self-Attn  & 40 & 3.49  & 10.92 & \textbf{--0.72} \\
                   & 50 & 3.29  &  9.26 & \textbf{--2.38} \\
                   & 60 & 3.43  & 10.21 & \textbf{--1.43} \\
    Dec Cross-Attn & 10 & 3.42  & 11.19 & \textbf{--0.45} \\
    Dec FFNs       & 30 & 3.95  & 12.16 & +0.52           \\
                   & 40 & 32.24 & 52.63 & +40.99          \\
    \midrule
    LayerNorms     & 10 & 3.40  & 11.01 & \textbf{--0.63} \\
    Bias           & 10 & 4.01  & 19.17 & +7.53           \\
    Conv Layers    & 50 & 3.42  & 11.64 & \textbf{+0.00}  \\
    Positional Emb & 30 & 3.68  & 13.28 & +1.64           \\
    Token Emb      & 30 & 3.85  & 13.48 & +1.84           \\
    Output Proj    & 20 & 3.44  & 10.54 & \textbf{--1.10} \\
    \bottomrule
  \end{tabular}
\end{table}

\subsection{Cross-dataset validation}
\label{ssec:validation}

To verify that regularization gains generalize beyond LibriSpeech, we evaluate pruning
configurations (identified using LibriSpeech sensitivity analysis) on Common Voice v15
English and TED-LIUM~3 \emph{without any mask recomputation or fine-tuning}. This
strict protocol ensures we assess true cross-corpus generalization.
Table~\ref{tab:cross_dataset} shows that improvements persist across acoustically
diverse conditions: accented crowd-sourced speech (Common Voice) and reverberant
conference presentations (TED-LIUM). Decoder self-attention at 50\% sparsity
consistently provides the largest WER reductions, while late encoder and output
projection pruning deliver smaller yet reliable gains. This suggests that the
identified redundancies are not dataset-specific but reflect architectural
over-parameterization that manifests across diverse acoustic environments.

\begin{table}[th]
  \caption{Cross-dataset validation of LibriSpeech configurations.}
  \label{tab:cross_dataset}
  \centering
  \footnotesize
  \begin{tabular}{@{}ll@{\hspace{0.4em}}c@{\hspace{0.4em}}c@{\hspace{0.4em}}c@{}}
    \toprule
    Dataset      & Component        & Sparsity (\%) & WER (\%) & $\Delta$WER     \\
    \midrule
    \emph{TED-3} & Baseline         &  0 & 6.43          & --              \\
                 & Late enc.\ (9--12) & 50 & 5.86        & \textbf{--0.57} \\
                 & Dec.\ self-attn  & 50 & \textbf{5.79} & \textbf{--0.64} \\
                 & Output proj.     & 20 & 6.48          & +0.05           \\
    \midrule
    \emph{CV-15} & Baseline         &  0 & 20.99          & --             \\
    (EN)         & Late enc.\ (9--12) & 50 & 19.78        & \textbf{--1.21} \\
                 & Dec.\ self-attn  & 50 & \textbf{18.45} & \textbf{--2.54} \\
                 & Output proj.     & 20 & 20.11          & \textbf{--0.88} \\
    \bottomrule
  \end{tabular}
\end{table}

\subsection{Sensitivity-aware one-shot compression}
\label{ssec:compression}

Beyond regularization, our sensitivity analysis enables aggressive one-shot compression
by allocating sparsity to resilient components while preserving fragile ones. As shown
in Section~\ref{ssec:encoder-decoder}, global magnitude pruning fails catastrophically
at 40\% sparsity (61.32\% WER). In contrast, component-specific allocation guided by
our sensitivity diagnostics achieves the same overall sparsity (40.8\%) while
preserving accuracy.

Table~\ref{tab:model_comparison} compares unpruned \Whisper{}-small against our
sensitivity-aware compressed variant. Despite reducing parameters from 241M to 143M and
compute from 4.55 to 2.77~GFLOPs, \textit{test-other} WER only worsens by 0.20\%,
while CER actually \emph{improves} by 1.06\%. This demonstrates that understanding
\emph{where} to prune is as critical as \emph{how much} to prune.

\begin{table}[th]
  \caption{Sensitivity-aware one-shot compression on LibriSpeech \textit{test-other}:
    \Whisper{}-small vs.\ Custom Pruned.}
  \label{tab:model_comparison}
  \centering
  \begin{tabular}{lrr}
    \toprule
                         & \Whisper{}-small & Custom Pruned \\
    \midrule
    WER (\%)             & 11.64  & 11.84  \\
    CER (\%)             &  6.97  &  5.91  \\
    Total parameters (M) & 241.73 & 143.11 \\
    Sparsity (\%)        &  0     & 40.8   \\
    GFLOPs               &  4.55  &  2.77  \\
    Sparse size (MB)     & 922.3  & 545.9  \\
    \bottomrule
  \end{tabular}
\end{table}

For reproducibility, we used the following sparsity thresholds: Encoder:
Conv.\ Layers~20\%, Self-Attn.\ 40\%, FFNs~55\%. Decoder: Self-Attn.\ 50\%,
Cross-Attn.\ 45\%, FFNs: early~25\%, middle~45\%, late~30\%, Token
Emb.\ 25\%, Output Proj.\ 25\%.

\section{Conclusion}
\label{sec:conclusion}

We have demonstrated that one-shot magnitude pruning can act as an implicit regularizer
for encoder-decoder ASR Transformers, improving generalization without fine-tuning and
challenging the view of pruning as purely a compression technique. Combining gradient-based and Fisher-based sensitivity diagnostics with component-wise experiments, we identify actionable asymmetries in \Whisper{}-small: pruning 50\% of decoder self-attention improves LibriSpeech \textit{test-other} WER by 2.38\% absolute (20.44\% relative), while pruning the late encoder block (layers 9--12) at 50\% yields a 1.72\% improvement. These gains transfer to Common Voice and TED-LIUM without mask recomputation. Beyond regularization,
sensitivity-aware sparsity allocation enables one-shot compression to 40.8\% sparsity where naive global pruning collapses, providing a reproducible methodology for deciding \emph{where} to prune in Transformer-based ASR.

\textbf{Limitations and future work.} Results are currently scoped to \Whisper{}-small on
English, an intentional choice to isolate the pruning-as-regularization effect in a
controlled setting. Pilot runs on other ASR architectures (Parakeet, wav2vec 2.0, Conformer) revealed each model's distinct component-level pruning sensitivities, indicating the framework's generality. Full empirical validation across these models is ongoing. Extending to multilingual scenarios, where language-specific representations may shift component-level resilience, and studying the interaction between sensitivity-guided pruning and fine-tuning are promising directions. Encoder-only CTC systems such as HuBERT-based ASR lack an autoregressive decoder, so the encoder--decoder asymmetry finding does not directly transfer, while the same sensitivity framework remains applicable.

\section{Generative AI Use Disclosure}

Generative AI tools were used only for language polishing (grammar and clarity). All scientific content, experimental results, and conclusions were produced by the authors.

\bibliographystyle{IEEEtran}
\bibliography{refs}

@article{srivastava2014dropout,
    author  = {Srivastava, Nitish and Hinton, Geoffrey and Krizhevsky, Alex
               and Sutskever, Ilya and Salakhutdinov, Ruslan},
    title   = {Dropout: A Simple Way to Prevent Neural Networks from Overfitting},
    journal = {Journal of Machine Learning Research},
    volume  = {15},
    pages   = {1929--1958},
    year    = {2014},
}

@inproceedings{krogh1991simple,
    author    = {Krogh, Anders and Hertz, John A.},
    title     = {A Simple Weight Decay Can Improve Generalization},
    booktitle = {Proc. Advances in Neural Information Processing Systems (NIPS)},
    pages     = {950--957},
    year      = {1991},
}

@inproceedings{ioffe2015batch,
    author    = {Ioffe, Sergey and Szegedy, Christian},
    title     = {Batch Normalization: Accelerating Deep Network Training by
                 Reducing Internal Covariate Shift},
    booktitle = {Proc. International Conference on Machine Learning (ICML)},
    pages     = {448--456},
    year      = {2015},
}

@article{tibshirani1996regression,
    author  = {Tibshirani, Robert},
    title   = {Regression Shrinkage and Selection via the Lasso},
    journal = {Journal of the Royal Statistical Society, Series B},
    volume  = {58},
    number  = {1},
    pages   = {267--288},
    year    = {1996},
}

@inproceedings{lecun1989optimal,
    author    = {LeCun, Yann and Denker, John S. and Solla, Sara A.},
    title     = {Optimal Brain Damage},
    booktitle = {Proc. Advances in Neural Information Processing Systems (NeurIPS)},
    pages     = {598--605},
    year      = {1989},
}

@inproceedings{hassibi1993optimal,
    author    = {Hassibi, Babak and Stork, David G. and Wolff, Gregory J.},
    title     = {Optimal Brain Surgeon and General Network Pruning},
    booktitle = {Proc. {IEEE} International Conference on Neural Networks},
    year      = {1993},
}

@inproceedings{han2015learning,
    author    = {Han, Song and Pool, Jeff and Tran, John and Dally, William J.},
    title     = {Learning both Weights and Connections for Efficient Neural Networks},
    booktitle = {Proc. Advances in Neural Information Processing Systems (NeurIPS)},
    pages     = {1135--1143},
    year      = {2015},
}

@inproceedings{voita2019analyzing,
    author    = {Voita, Elena and Talbot, David and Moiseev, Fedor
                 and Sennrich, Rico and Titov, Ivan},
    title     = {Analyzing Multi-Head Self-Attention: Specialized Heads Do the
                 Heavy Lifting, the Rest Can Be Pruned},
    booktitle = {Proc. Annual Meeting of the Association for Computational
                 Linguistics (ACL)},
    pages     = {5797--5808},
    year      = {2019},
}

@inproceedings{michel2019heads,
    author    = {Michel, Paul and Levy, Omer and Neubig, Graham},
    title     = {Are Sixteen Heads Really Better than One?},
    booktitle = {Proc. Advances in Neural Information Processing Systems (NeurIPS)},
    pages     = {14014--14024},
    year      = {2019},
}

@inproceedings{behnke2020losing,
    author    = {Behnke, Maximiliana and Heafield, Kenneth},
    title     = {Losing Heads in the Lottery: Pruning Transformer Attention in
                 Neural Machine Translation},
    booktitle = {Proc. Conference on Empirical Methods in Natural Language
                 Processing (EMNLP)},
    pages     = {2664--2674},
    year      = {2020},
}

@article{gale2019state,
    author  = {Gale, Trevor and Elsen, Erich and Hooker, Sara},
    title   = {The State of Sparsity in Deep Neural Networks},
    journal = {arXiv preprint arXiv:1902.09574},
    year    = {2019},
}

@inproceedings{bartoldson2020generalization,
    author    = {Bartoldson, Brian and Morcos, Ari S. and Barbu, Adrian
                 and Erlebacher, Gordon},
    title     = {The Generalization-Stability Tradeoff in Neural Network Pruning},
    booktitle = {Proc. Advances in Neural Information Processing Systems (NeurIPS)},
    pages     = {20852--20864},
    year      = {2020},
}

@inproceedings{jin2022pruning,
    author    = {Jin, Tian and Carbin, Michael and Roy, Daniel M.
                 and Frankle, Jonathan and Dziugaite, Gintare Karolina},
    title     = {Pruning's Effect on Generalization Through the Lens of Training
                 and Regularization},
    booktitle = {Proc. Advances in Neural Information Processing Systems (NeurIPS)},
    year      = {2022},
}

@inproceedings{frankle2019lottery,
    author    = {Frankle, Jonathan and Carbin, Michael},
    title     = {The Lottery Ticket Hypothesis: Finding Sparse, Trainable
                 Neural Networks},
    booktitle = {Proc. International Conference on Learning Representations (ICLR)},
    year      = {2019},
}

@inproceedings{lee2019snip,
    author    = {Lee, Namhoon and Ajanthan, Thalaiyasingam and Torr, Philip H.S.},
    title     = {{SNIP}: Single-Shot Network Pruning Based on Connection Sensitivity},
    booktitle = {Proc. International Conference on Learning Representations (ICLR)},
    year      = {2019},
}

@inproceedings{wang2020picking,
    author    = {Wang, Chaoqi and Zhang, Guodong and Grosse, Roger},
    title     = {Picking Winning Tickets Before Training by Preserving Gradient Flow},
    booktitle = {Proc. International Conference on Learning Representations (ICLR)},
    year      = {2020},
}

@inproceedings{tanaka2020pruning,
    author    = {Tanaka, Hidenori and Kunin, Daniel and Yamins, Daniel L.K.
                 and Ganguli, Surya},
    title     = {Pruning Neural Networks Without Any Data by Iteratively
                 Conserving Synaptic Flow},
    booktitle = {Proc. Advances in Neural Information Processing Systems (NeurIPS)},
    year      = {2020},
}

@inproceedings{singh2020woodfisher,
    author    = {Singh, Sidak Pal and Alistarh, Dan},
    title     = {{WoodFisher}: Efficient Second-Order Approximation for Neural
                 Network Compression},
    booktitle = {Proc. Advances in Neural Information Processing Systems (NeurIPS)},
    pages     = {18098--18109},
    year      = {2020},
}

@inproceedings{liu2021group,
    author    = {Liu, Liyang and Zhang, Shilong and Kuang, Zhanghui and Zhou, Aojun
                 and Xue, Jing-Hao and Wang, Xinjiang and Chen, Yimin
                 and Yang, Wenming and Liao, Qingmin and Zhang, Wayne},
    title     = {Group Fisher Pruning for Practical Network Compression},
    booktitle = {Proc. International Conference on Machine Learning (ICML)},
    pages     = {7021--7032},
    year      = {2021},
}

@inproceedings{kwon2022fast,
    author    = {Kwon, Woosuk and Kim, Sehoon and Mahoney, Michael W.
                 and Hassoun, Joseph and Keutzer, Kurt and Gholami, Amir},
    title     = {A Fast Post-Training Pruning Framework for Transformers},
    booktitle = {Proc. Advances in Neural Information Processing Systems (NeurIPS)},
    year      = {2022},
}

@inproceedings{peng2023structured,
    author    = {Peng, Yifan and Kim, Kwangyoun and Wu, Felix and Sridhar, Prashant
                 and Watanabe, Shinji},
    title     = {Structured Pruning of Self-Supervised Pre-Trained Models for Speech
                 Recognition and Understanding},
    booktitle = {Proc. {IEEE} International Conference on Acoustics, Speech and
                 Signal Processing (ICASSP)},
    year      = {2023},
}

@inproceedings{kim2023adapt,
    author    = {Kim, Hyeon Soo and Cho, Chung Hyeon and Won, Hyejin and Park, Kyung Ho},
    title     = {Adapt and Prune Strategy for Multilingual Speech Foundational Model
                 on Low-Resourced Languages},
    booktitle = {Proc. 3rd Workshop on Multi-lingual Representation Learning (MRL)},
    pages     = {85--94},
    year      = {2023},
}

@inproceedings{gu2024sparseWAV,
    author    = {Gu, Tianteng and Liu, Bei and Shao, Hao and Qian, Yanmin},
    title     = {{SparseWAV}: Fast and Accurate One-Shot Unstructured Pruning for
                 Large Speech Foundation Models},
    booktitle = {Proc. {INTERSPEECH} 2024 -- 25\textsuperscript{th} Annual Conference
                 of the International Speech Communication Association},
    year      = {2024},
}

@article{kirkpatrick2016overcoming,
    author  = {Kirkpatrick, James and Pascanu, Razvan and Rabinowitz, Neil
               and Veness, Joel and Desjardins, Guillaume and Rusu, Andrei A.
               and Milan, Kieran and Quan, John and Ramalho, Tiago
               and Grabska-Barwinska, Agnieszka and Hassabis, Demis
               and Clopath, Claudia and Kumaran, Dharshan and Hadsell, Raia},
    title   = {Overcoming Catastrophic Forgetting in Neural Networks},
    journal = {Proceedings of the National Academy of Sciences},
    volume  = {114},
    number  = {13},
    pages   = {3521--3526},
    year    = {2017},
}

@inproceedings{radford2022whisper,
    author    = {Radford, Alec and Kim, Jong Wook and Xu, Tao and Brockman, Greg
                 and McLeavey, Christine and Sutskever, Ilya},
    title     = {Robust Speech Recognition via Large-Scale Weak Supervision},
    booktitle = {Proc. International Conference on Machine Learning (ICML)},
    pages     = {28492--28518},
    year      = {2023},
}

@inproceedings{panayotov2015librispeech,
    author    = {Panayotov, Vassil and Chen, Guoguo and Povey, Daniel
                 and Khudanpur, Sanjeev},
    title     = {{LibriSpeech}: An {ASR} Corpus Based on Public Domain Audio Books},
    booktitle = {Proc. {IEEE} International Conference on Acoustics, Speech and
                 Signal Processing (ICASSP)},
    pages     = {5206--5210},
    year      = {2015},
}

@inproceedings{ardila2020common,
    author    = {Ardila, Rosana and Branson, Megan and Davis, Kelly
                 and Henretty, Michael and Kohler, Michael and Meyer, Josh
                 and Morais, Reuben and Saunders, Lindsay and Tyers, Francis M.
                 and Weber, Gregor},
    title     = {Common Voice: A Massively-Multilingual Speech Corpus},
    booktitle = {Proc. International Conference on Language Resources and
                 Evaluation (LREC)},
    year      = {2020},
}

@inproceedings{hernandez2018ted,
    author    = {Hernandez, Fran\c{c}ois and Nguyen, Vincent and Ghannay, Sahar
                 and Tomashenko, Natalia and Est\`eve, Yannick},
    title     = {{TED-LIUM} 3: Twice as Much Data and Corpus Repartition for
                 Experiments on Speaker Adaptation},
    booktitle = {Proc. International Conference on Speech and Computer (SPECOM)},
    pages     = {198--208},
    year      = {2018},
}

\end{document}